\magnification 1200
\centerline {\bf New Structures in the Theory of the Laser Model
II:} 
\vskip 0.3cm
\centerline {\bf Microscopic Dynamics and a Non-Equilibrium Entropy Principle}
\vskip 1cm
\centerline {\bf by Fabio Bagarello$^{a)}$ and Geoffrey L.
Sewell$^{b)}$}
\vskip 0.5cm
\centerline {\bf Department of Physics, Queen Mary and Westfield
College}
\vskip 0.3cm
\centerline {\bf Mile End Road, London E1 4NS, England} 
\vskip 2cm
\centerline {\bf Abstract.}
\vskip 0.5cm
In a recent article, Alli and Sewell formulated a new version of
the Dicke-Hepp-Lieb laser model in terms of quantum dynamical 
semigroups, and thereby extended the macroscopic picture of the
model. In the present article, we complement that picture with
a corresponding microscopic one, which carries the following new
results. (a) The local microscopic dynamics of the model is
piloted by the classical, macroscopic field, generated by the
collective action of its components; (b) the global state of the
system carries no correlations between its constituent atoms
after transient effects have died out; and (c) in the latter
situation, the state of the system at any time $t$ maximises its
entropy density, subject to the constraints imposed by the
instantaneous values of its macroscopic variables.
\vfill\eject
\centerline {\bf I. Introduction}
\vskip 0.3cm
The classic work by Hepp and Lieb$^{(1)}$ (HL) on the theory of
the Dicke laser model$^{(2)}$ represents a landmark in the
constructive quantum theory of phase transitions, far from
equilibrium, in open quantum systems. Most notably, it
demonstrated how the model undergoes a transition from normal to
coherent radiation when the optical pumping strength reaches a
critical value. This substantiated ideas that had been proposed
by Graham and Haken$^{(3)}$ at a more heuristic level. 
\vskip 0.2cm 
In a recent article, which we shall refer to as (I), Alli and
Sewell$^{(4)}$ constructed a new version$^{(5)}$ of the Dicke
model, in which the dynamics of the matter and radiation is
governed by a quantum dynamical semigroup, which incorporates the
dissipative action of an array of pumps and sinks: by contrast,
the HL model represented a conservative system, comprising the
matter, radiation and these reservoirs.
The principle new results of (I) were (a) a generalisation of the
HL form of the dynamics of the matter-radiation composite, (b)
a demonstration that the model exhibits optically chaotic phases,
in addition to the normal and coherent ones, and (c) a
generalisation of the theory of quantum dynamical
semigroups$^{(6)}$ to ones with unbounded generators.
\vskip 0.2cm
The theories of both HL and (I) were centred on the macroscopic
equations of motion, derived from the underlying quantum dynamics
of their respective models in a limit where their sizes become
infinite. In fact, they did not provide any treatment of the
microdynamics in this limit. The object of the present article
is to complement the macroscopic picture of (I) with a
corresponding microscopic one.
\vskip 0.2cm
In seeking a microscopic-cum-macroscopic picture of the dynamics
of the model, we note that this has already been achieved for
certain {\it conservative} systems with long range forces. For
such systems, Morchio and Strocchi$^{(7)}$ have shown that
the evolutes, $b_{t},$ of the local microscopic observables, $b,$
depend not only on $b$ and the time $t,$ but also on certain
macroscopic observables, i.e. 'observables at infinity': this
represents a picture completely different from that of the Haag-
Kastler scheme$^{(8)}$, in which the dynamics of the micro-
observables is autonomous. Furthermore, the explicit structure
of the equations of motion for the microscopic and
macroscopic observables, $b_{t}$ and $B_{t},$ has been obtained
by Bagarello and Morchio$^{(9)}$ (BM) for a class of
conservative, mean field theoretic models in the following form.
$${db_{t}\over dt}=f(b_{t},B_{t}); \ {dB_{t}\over dt}=F(B_{t}).
\eqno(1.1)$$
Thus, the microscopic dynamics is piloted by the macroscopic
observables, which, for their part, evolve according to a
self-contained law.
\vskip 0.2cm
The question now arises as to whether a similar state of affairs
prevails in the case of open dissipative systems, such as the
laser model; and the first thing to say about that is that the
methods of BM, which led to equn. (1.1), are not applicable
there. The reason is simply that those methods depended heavily
on the fact that, in a conservative system, the dynamics is
automorphic and so preserves the algebraic structure of the
observables. In an open dissipative system, on the other hand,
the dynamics is governed by a semigroup$^{(4)}$ of
transformations
of the observables that does not preserve that structure, i.e.,
in general, $(ab)_{t}{\neq}a_{t}b_{t}.$
\vskip 0.2cm
Thus, any attempt to derive equations of the form (1.1) for
the laser model must be based on methods that are applicable to
dissipative systems. Here, the essential problem is to pass to
a limit where the size of the system becomes infinite, since
the dynamics of the finite version of the model has already been 
formulated in (I). In fact, we surmount this problem by means of
a compactness argument, that lends itself to the application of
the Arzela-Ascoli theorem. As a result, we do indeed obtain a
microscopic-cum-macroscopic dynamics of the form (1.1), though
now, of course, the functions $f$ and $F$ incorporate the
dissipative action of the pumps and sinks on the system.
\vskip 0.2cm
The principal new results that we obtain in this way are of a
very simple character, and may be summarised as follows.
\vskip 0.2cm
(1) The microscopic observables of {\it this particular model}
are those of the matter only, since the radiation field reduces
to a classical macroscopic one.
\vskip 0.2cm
(2) The dynamics of the local microscopic observables is piloted
by that classical radiation field.
\vskip 0.2cm
(3) The microdynamics contains certain transients, whose
lifetimes are of the order of that of an atomic excited state.
This dynamics simplifies considerably after these transients have
died out, and leads to the decorrelation of the observables of
different atoms. Furthermore, the resultant global microstate of
the system is stationary and space-translationally invariant in
the normal phase, and periodic with respect to both space and
time in the coherent one. Evidently, it is very complicated in
the chaotic phase.
\vskip 0.2cm
(4) Following the decay of the transients, the instantaneous
state of the model is determined by a non-equilibrium variational
principle. Specifically, this state maximises the global entropy 
density, subject to the constraints imposed by the prevailing
values of the macroscopic observables. As far as we know, this is 
the first example for which a principle of this kind has been 
proved for the non-equilibrium regime. 
\vskip 0.2cm
We shall present our arguments as follows. In Sections II and
III, we shall provide a brief summary of (I), re-expressed in a
form suitable for our present purposes. Specifically, Section II
will be devoted to a formulation of the model and Section III to
a resume of its macroscopic properties. In Section IV, we shall
derive the above results (1) and (2) by passing to the infinite
size limit of the quantum dynamics of the finite version of the
model. We shall then derive the results (3) and (4) in Sections
V and VI, respectively. We shall conclude, in Section VII, with
a brief discussion of our main results and their possible
ramifications.  
\vskip 0.5cm
\centerline {\bf II. The Model.}
\vskip 0.3cm
Our model, as formulated in (I), is a dissipative quantum system,
${\Sigma}^{(N)},$ consisting of a chain of $2N+1$ identical
two-level atoms interacting with an $n-$mode radiation field. We
label it
by $N$ only, since we shall be concerned with its properties in
a limit
where $N$ tends to infinity, while $n$ remains fixed and finite.
We formulate ${\Sigma}^{(N)}$ as a quantum dynamical system
$({\cal B}^{(N)},T^{(N)},{\cal S}^{(N)}),$ where ${\cal B}^{(N)}$
is a ${\star}-$algebra of observables, $T^{(N)}$ is a
one-parameter semigroup of completely positive (CP)
transformations of ${\cal B}^{(N)},$ and ${\cal S}^{(N)}$ is a
family of normal states on ${\cal B}^{(N)},$ that is stable under
the dual of $T^{(N)}.$ Most importantly, $T^{(N)}$ incorporates
the dissipative action of pumps and sinks, which do not appear
explicitly in the model, on the matter-radiation system. We build
the model from its constituent parts as follows.
\vskip 0.2cm
{\bf The Single Atom.} This is assumed to be a two-state atom or
spin, ${\Sigma}_{at}.$ Its algebra of observables, ${\cal
A}_{at},$ is that of the two-by-two matrices, and is thus the
linear span of Pauli matrices
$({\sigma}_{x},{\sigma}_{y},{\sigma}_{z})$ and the identity,
$I.$ Its algebraic structure is provided by the relations
$${\sigma}_{x}^{2}={\sigma}_{y}^{2}={\sigma}_{z}^{2}=I;
\ {\sigma}_{x}{\sigma}_{y}=i{\sigma}_{z}, \
etc.\eqno(2.1)$$
We define the spin raising and lowering operators
$${\sigma}_{{\pm}}={1\over 2}
({\sigma}_{x}{\pm}i{\sigma}_{y}).\eqno(2.2)$$ 
We assume that the atom is coupled to a pump and a sink, and
that accordingly (cf. (I)) its dynamics is given
by a one-parameter semigroup ${\lbrace}T_{at}(t){\vert}t
{\in}{\bf R}_{+}{\rbrace}$ of completely
positive, identity preserving contractions of ${\cal A}_{at},$
whose generator, $L_{at},$ is of the following form.
$$L_{at}{\sigma}_{\pm}=-
({\gamma}_{1}{\mp}i{\epsilon}){\sigma}_{\pm}; \
L_{at}{\sigma}_{z}=-{\gamma}_{2}({\sigma}_{z}-
{\eta}I),\eqno(2.3)$$
where ${\epsilon}(>0)$ is the energy difference between the
ground and excited states of an atom, and the ${\gamma}$'s and
${\eta}$ are constants whose values are determined by the atomic
coupling to the energy source and sink, and are subject to the
restrictions that
$$0<{\gamma}_{2}{\leq}2{\gamma}_{1}; 
\ -1{\leq}{\eta}{\leq}1.\eqno(2.4)$$
In particular, ${\eta}$ is positive or negative according to
whether the coupling of the atom to the pump or the sink is the
stronger. In the former case, these couplings drive the atom to
a terminal mixed state with {\it inverted population}, i.e. with
higher occupation probability for the excited state than for the
ground state.
\vskip 0.3cm
{\bf The Matter.} This consists of $2N+1$ non-interacting copies
of ${\Sigma}_{at},$ located at the sites $r=-N,. \ .,N$ of the
one-dimensional lattice ${\bf Z}.$ Thus, at each site, $r,$ there
is a copy, ${\Sigma}_{r},$ of ${\Sigma}_{at},$ whose algebra
of observables, ${\cal A}_{r},$ and dynamical semigroup, $T_{r},$
are isomorphic with ${\cal A}_{at}$ and $T_{at},$ respectively.
We denote by ${\sigma}_{r,u}$ the copy of
${\sigma}_{u}$ at $r,$ for $u=x,y,z,{\pm}.$ 
\vskip 0.2cm
We define the algebra of observables, ${\cal A}^{(N)},$ and 
the dynamical semigroup, $T_{mat}^{(N)},$
of the matter to be ${\otimes}_{r=-N}^{N}{\cal A}_{r}$ and
${\otimes}_{r=-N}^{N}T_{r},$ respectively. Thus, ${\cal
A}^{(N)}$ is the algebra of linear transformations of ${\bf
C}^{4N+2}.$ We identify elements $A_{r}$ of ${\cal A}_{r}$
with those of ${\cal A}^{(N)}$ given by their tensor
products with the identity operators attached to the remaining
sites. Under this identification, the commutant, ${\cal
A}_{r}^{\prime},$ of ${\cal A}_{r}$ is the tensor product
${\otimes}_{s{\neq}r}{\cal A}_{s}.$ 
\vskip 0.2cm
It follows from these specifications that the generator,
$L_{mat}^{(N)},$ of $T_{mat}^{(N)}$ is given by the formula
$$L_{mat}^{(N)}={\sum}_{r=-N}^{N}L_{r},\eqno(2.5)$$
where  
$$L_{r}{\sigma}_{r,{\pm}}=-
({\gamma}_{1}{\mp}i{\epsilon}){\sigma}_{r,{\pm}}; \
L_{r}{\sigma}_{r,z}=-{\gamma}_{2}({\sigma}_{r,z}-
{\eta}I);$$
$$and \ L_{r}(A_{r}A_{r}^{\prime})=(L_{r}A_{r})A_{r}^{\prime} 
 \ {\forall}A_{r}{\in}{\cal A}_{r}, \ 
A_{r}^{\prime}{\in}{\cal A}_{r}^{\prime}\eqno(2.6)$$
\vskip 0.3cm
{\bf The Radiation.} We assume that the radiation field
consists of $n(<{\infty})$ modes, represented by creation and
destruction operators ${\lbrace}a_{l}^{\star},a_{l}{\vert}l=0,.
\ .,n-1{\rbrace}$ in a Fock-Hilbert space ${\cal H}_{rad},$ 
as defined by the standard specifications that (a) these
operators satisfy the CCR,
$$[a_{l},a_{m}^{\star}]={\delta}_{lm}I; \
[a_{l},a_{m}]=0,\eqno(2.7)$$
and (b) ${\cal H}_{rad}$ contains a (vacuum) vector ${\Phi}$,
that is annihilated by each of the $a$'s and is cyclic w.r.t. the
algebra of polynomials in the $a^{\star}$'s.
\vskip 0.2cm
We define the Weyl map $z=(z_{0},. \
.,z_{n-1}){\rightarrow}W(z),$ of $C^{n}$ into
${\cal L}({\cal H}_{rad})$ by the formula
$$W(z){\equiv}W(z_{0},. \ .,z_{n-1})=
{\exp}[i(z.a+(z.a)^{\star})], \ with \ 
z.a={\sum}_{0}^{n-1}z_{l}a_{l}\eqno(2.8)$$
We then define ${\cal R}$ to be the $^{\star}-$algebra of
polynomials in the $a$'s, $a^{\star}$'s and the Weyl operators
$W(z),$ with $z$ running through ${\bf C}^{n}.$ Thus, in view of
the CCR (2.7), this algebra is the linear span of terms of the
form $W(z)a_{l_{1}}^{\sharp}. \ .a_{l_{m}}^{\sharp},$ where
$a^{\sharp}$ denotes $a$ or $a^{\star}.$ Equivalently, it is the
linear span of the derivatives, of all orders, of the operators
$W(t_{0}z_{0},. \ .,t_{n-1}z_{n-1})$ w.r.t. the real variables
$(t_{0},. \ .,t_{n-1}).$
\vskip 0.2cm
We assume that the radiation dynamics is given by the canonical
extension to ${\cal R}$ of Vanheuverszwijn semigroup$^{(10)}, \
T_{rad},$ of quasi-free CP transformations of the Weyl algebra
of linear combinations of ${\lbrace}W(z){\vert}z{\in}{\bf
C}^{n}{\rbrace}.$ The formal generator of this semigroup is
$$L_{rad}={\sum}_{l=0}^{n-1}\bigl(i{\omega}_{l}
[a_{l}^{\star}a_{l},.]+2{\kappa}_{l}a_{l}^{\star}(.)a_{l}-
{\kappa}_{l}{\lbrace}a_{l}^{\star}a_{l},.{\rbrace}\bigr),\eqno
(2.9)$$
where ${\lbrace}.,.{\rbrace}$ denotes anticommutator, and the
frequencies, ${\omega}_{l},$ and the damping constants,
${\kappa}_{l},$ are positive.
\vskip 0.3cm
{\bf The Composite System.} This is the coupled system,
${\Sigma}^{(N)},$ comprising the matter and radiation. We
assume that its algebra of observables, ${\cal B}^{(N)}$, is the
tensor product ${\cal A}^{(N)}{\otimes}{\cal R}.$
Thus, ${\cal B}^{(N)},$ like ${\cal R},$ is an algebra
of both bounded and unbounded operators in the Hilbert space
${\cal H}^{(N)}:={\bf C}^{4N+2}{\otimes}{\cal H}_{rad}$. We shall
identify elements $A, \ R,$ of ${\cal A}^{(N)}, \ {\cal R},$ with
$A{\otimes}I_{rad}$ and $I_{mat}{\otimes}R$, respectively. 
\vskip 0.2cm
We assume that the matter-radiation coupling is dipolar and is
given by the interaction Hamiltonian
$$H_{int}^{(N)}=i(2N+1)^{-1/2}{\sum}_{r=-N}^{N}{\sum}_{l=0}^{n-1}
{\lambda}_{l}({\sigma}_{r,-}a_{l}^{\star}{\exp}(-2{\pi}ilr/n)-
h.c.),\eqno(2.10)$$
where the ${\lambda}$'s are real-valued, $N-$independent coupling
constants. Correspondingly, we define the radiation field,
${\phi}^{(N)},$ of the model by the stipulation that its value
at the site $r$ is the coefficient of ${\sigma}_{r,+}$ in this
formula. Thus,
$${\phi}_{r}^{(N)}=-i(2N+1)^{-1/2}{\sum}_{l=0}^{n-1}{\lambda}_{l}
a_{l}{\exp}(2{\pi}ilr/n).\eqno(2.11)$$ 
\vskip 0.2cm
We now need to define the state space, ${\cal S}^{(N)},$ and the
dynamical semigroup, $T^{(N)},$ in a way that takes account of
the unboundedness both of $H_{int}^{(N)}$ and of some of the
elements of ${\cal B}^{(N)}.$ To this end, we start by defining
${\cal D}_{0}^{(N)}$ to be the space of density matrices in
${\cal H}^{(N)},$ with the trace norm topology, and ${\cal
D}_{1}^{(N)}$ to be the subset of its elements, ${\rho}^{(N)},$
for which $Tr(B^{\star}{\rho}^{(N)}B)$ is finite for all $B$ in
${\cal B}^{(N)}.$ We define ${\cal S}_{1}^{(N)}$ to be the set
of positive, normalised, linear functionals, ${\psi}^{(N)},$ on
${\cal B}^{(N)},$ that are in the one-to-one correspondence with
the ${\cal D}_{1}^{(N)}-$class density matrices, given by the
standard formula ${\psi}^{(N)}(B)=Tr({\rho}^{(N)}B).$ We then
adopt the formulation of $T^{(N)}$ and ${\cal S}^{(N)}$ provided
by (I), in which $T^{(N)}$ is constructed as the modification
of $T_{mat}^{(N)}{\otimes}T_{rad}$ due to the interaction
$H_{int}^{(N)},$ and ${\cal S}^{(N)}$ is a 'large' subset of
${\cal S}_{1}^{(N)}$ that is stable w.r.t. this dynamical
semigroup. Specifically, the construction yields the following
results$^{(4)}$. 
\vskip 0.2cm
(1) The density matrices corresponding to ${\cal S}^{(N)}$ form
a dense subset of ${\cal D}_{0}^{(N)},$ in the topology
corresponding to the trace norm.
\vskip 0.2cm
(2) ${\cal S}^{(N)}$ is stable under the transformations
${\psi}^{(N)}{\rightarrow}{\psi}^{(N)}{\circ}T^{(N)}(t):={\psi
}_{t}^{(N)}.$
\vskip 0.2cm
(3) $${d\over dt}{\psi}_{t}^{(N)}(B)={\psi}_{t}^{(N)}(L^{(N)}B)
\ {\forall}B{\in}{\cal B}^{(N)}, \ t{\in}{\bf
R}_{+},\eqno(2.12)$$
where
$$L^{(N)}=L_{mat}^{(N)}+L_{rad}+i[H_{int}^{(N)},.].\eqno(2.13)$$
This completes our specification of ${\Sigma}^{(N)}=({\cal
B}^{(N)},T^{(N)},{\cal S}^{(N)}).$
\vskip 0.3cm
{\bf The Sequence of Systems ${\lbrace}{\Sigma}^{(N)}{\rbrace}.$}
Since we shall be concerned with the properties of the above
model in a limit where $N$ tends to infinity and $n$ remains
fixed, we need to consider the sequence of systems
${\lbrace}{\Sigma}^{(N)}{\vert}N{\in}{\bf N}{\rbrace}.$
Specifically, we shall investigate the evolution of these systems
from initial states, ${\psi}^{(N)},$ for which the mean photon
number is not super-extensive, i.e., its ratio to $N$ is
bounded. Thus, we assume that, for some finite constant $C,$  
$${\psi}^{(N)}(a_{l}^{\star}a_{l})<CN, \ 
{\forall}N{\in}{\bf N},l{\in}[0,n-1].\eqno(2.14)$$
This exclusion of super-extensivity was shown in (I) to persist,
in a uniform way, at later times, in that
$${\psi}_{t}^{(N)}(a_{l}^{\star}a_{l})<DN, \ {\forall}
N{\in}{\bf N},t{\in}{\bf R}_{+},l{\in}[0,n-1],\eqno(2.15)$$
where $D$ is a finite constant.
\vskip 0.5cm
\centerline {\bf III. Macroscopic Dynamics and Phase Structure} 
\vskip 0.3cm
We formulate the macroscopic description of the model, as in (I),
in terms of the global intensive observables 
$$s_{l}^{(N)}=(2N+1)^{-1}{\sum}_{r=-N}^{N}
{\sigma}_{r,-}{\exp}(-2{\pi}ilr/n); \ l=0,.. \
.,n-1\eqno(3.1)$$
and
$$p_{l}^{(N)}=(2N+1)^{-1}{\sum}_{r=-N}^{N}
{\sigma}_{r,z}{\exp}(-2{\pi}ilr/n); \ l=0,.. \
.,n-1,\eqno(3.2)$$
together with the operators
$${\alpha}_{l}^{(N)}=(2N+1)^{-1/2}a_{l}; \ l=0,.. \
.,n-1,\eqno(3.3)$$
corresponding to a scaling of the number operators
$a_{l}^{\star}a_{l}$ in units of $2N+1$. We note that the set of
$p^{(N)}$'s is the same as that of their adjoints, since  
$$p_{0}^{(N){\star}}=p_{0}^{(N)}; \ and \ p_{l}^{(N){\star}}
=p_{n-l}^{(N)} \ for \ l=1,. \ .,n-1.\eqno(3.4)$$
The set 
$M^{(N)}:={\lbrace}s_{l}^{(N)},s_{l}^{(N){\star}},p_{l}^{(N)},
{\alpha}_{l}^{(N)},{\alpha}_{l}^{(N){\star}}{\rbrace}$ of
macroscopic variables is a Lie algebra w.r.t. commutation, and,
by equns. (2.10), (2.11) and (3.3), the radiation field and the
interaction Hamiltonian are the algebraic functions of them given
by
$${\phi}_{r}^{(N)}=-
i{\sum}_{l=0}^{n-1}{\lambda}_{l}{\alpha}_{l}^{(N)}
{\exp}(2{\pi}irl/n)\eqno(3.5)$$
and
$$H_{int}^{(N)}=i(2N+1){\sum}_{l=0}^{n-1}{\lambda}_{l}
(s_{l}^{(N)}{\alpha}_{l}^{(N){\star}}-h.c.).\eqno(3.6)$$
In (I), the dynamics of $M^{(N)}$ was extracted from the
microscopic equation of motion (2.12), in a limit where
$N{\rightarrow}{\infty}$ and $n$ remains fixed and finite,
subject to the assumptions of Section 2. As one might anticipate 
from the fact that the commutators of the elements of $M^{(N)}$
tend normwise to zero in this limit, this macrodynamics is
classical. 
\vskip 0.2cm
To describe this dynamics, we simplify our notation by 
defining $x^{(N)}{\equiv}(x_{1}^{(N)},. \ .,x_{5n}^{(N)})$ to be
the $5n$ self-adjoint operators, which, by equns. (3.1)-(3.4),
comprise the Hermitian and anti-Hermitian parts of the elements
of $M^{(N)}.$ Correspondingly, in anticipation of a classical
limit for the dynamics of these observables, as
$N{\rightarrow}{\infty},$ we introduce a phase space, $X={\bf
R}^{5n}$ and, for each point $x=(x_{1},. \ .,x_{5n}){\in}X,$ we
define ${\lbrace}{\alpha}_{l},s_{l},p_{l}{\vert}l=0,. \
.,n-1{\rbrace}$ to be the complex numbers related to $x$ 
in precisely the same way that the elements of $M^{(N)}$ are to
$x^{(N)}.$ We define ${\cal C}_{0}(X)$ to be the space
of bounded continuous functions on $X,$ that tend to zero at
infinity. 
\vskip 0.2cm
We formulate the macroscopic dynamics of ${\Sigma}^{(N)}$ in
terms of the time-dependent quantum characteristic function
${\mu}_{t}^{(N)}$ on $X,$ defined by the formula
$${\mu}_{t}^{(N)}(y)=
{\psi}_{t}^{(N)}\bigl({\exp}(iy.x^{(N)})\bigr)
\ {\forall}y{\in}X,t{\in}{\bf R}_{+}.\eqno(3.7)$$
Further (cf. (I)), we relate the limiting form of this function,
as $N{\rightarrow}{\infty},$ to the following classical dynamical
system.  
\vskip 0.2cm
{\bf Definition 3.1.} We define ${\cal K}$ to be the flow in $X,$
whose equation of motion, 
$${dx_{t}\over dt}=F(x_{t}),\eqno(3.8)$$
is of the explicit form 
$${d{\alpha}_{l,t}\over dt}=-
(i{\omega}_{l}+{\kappa}_{l}){\alpha}_{l,t}+
{\lambda}_{l}s_{l,t}\eqno(3.9a)$$
$${ds_{l,t}\over dt}=-(i{\epsilon}+{\gamma}_{1})s_{l,t}+
{\sum}_{m=0}^{n-1}{\lambda}_{m}p_{[l-m],t}{\alpha}_{m,t}
\eqno(3.9b)$$
and
$${dp_{l,t}\over dt}=-{\gamma}_{2}(p_{l,t}-{\eta}{\delta}_{l,0})
-2{\sum}_{m=0}^{n-1}{\lambda}_{m}({\overline {\alpha}}_{m,t}
s_{[l+m],t}+{\alpha}_{m,t}{\overline s}_{[m-l],t}),\eqno(3.9c)$$
where $[l{\pm}m]=l{\pm}m \ (mod \ n).$ 
\vskip 0.3cm
The following Propositions were proved in (I).
\vskip 0.3cm
{\bf Proposition 3.2.} {\it The equation of motion (3.8) (i.e.
(3.9)) has a unique global solution, corresponding to a one-
parameter semigroup of transformations
$x{\rightarrow}{\tau}_{t}x({\equiv}x_{t})$ of $X,$ that maps the
compacts into compacts.}
\vskip 0.3cm
{\bf Proposition 3.3.} {\it (i) Under the above assumptions, 
${\mu}_{t}^{(N)}$ tends pointwise, as $N{\rightarrow}{\infty},$
to the characteristic function, ${\mu}_{t},$ of a probability
measure, $m_{t},$ on $X,$ i.e.
$${\lim}_{N\to\infty}{\mu}_{t}^{(N)}(y)={\mu}_{t}(y)={\int}_{X}
{\exp}i(x.y)dm_{t}(x),\eqno(3.10)$$
convergence being uniform on the $X-$compacts.
\vskip 0.2cm
(ii) The evolution of $m_{t}$ is induced by the dynamical
semigroup, ${\tau},$ of ${\cal K},$ according to the formula
$${\int}_{X}dm_{t}f={\int}_{X}dm_{0}f{\circ}{\tau}(t) \ {\forall}
f{\in}{\cal C}_{0}(X), \ t{\in}{\bf R}_{+}.\eqno(3.11)$$
\vskip 0.2cm
(iii) In the particular case of a pure phase, where $m_{0}$ is
a Dirac measure, ${\delta}_{x},$ with support at some point
$x,$ then $m_{t}={\delta}_{x_{t}}.$} 
\vskip 0.3cm
{\bf Comment.} This last Proposition signifies that, in the limit
$N{\rightarrow}{\infty},$ the macroscopic variables
$({\alpha}^{(N)},s^{(N)},p^{(N)})$ reduce to classical ones
$({\alpha},s,p).$ Correspondingly, the radiative field,
${\phi}^{(N)},$ as given by equn. (3.5), reduces to a classical
one, ${\phi},$ whose value at the site $r$ and time $t$ is
$${\phi}_{r,t}=-i{\sum}_{l=0}^{n-1}{\lambda}_{l}
{\alpha}_{t,l}{\exp}(2{\pi}irl/n).\eqno(3.12)$$
\vskip 0.2cm
{\bf Phase Structure.} By Def. 3.1, Prop. 3.2 and Prop.
3.3(iii), the properties of the pure phase are governed by the
classical dynamical system ${\cal K}.$  A study of this system
yields the following results (cf. (I)).
\vskip 0.2cm
(1) For ${\eta}$ less than a certain specified critical value,
${\eta}_{1} \ (>0),$ the system has a unique stable stationary
state
corresponding to the fixed point, $x_{0}({\in}X),$ given by
$${\alpha}_{l}=0; \ s_{l}=0; \ p_{l}={\eta}{\delta}_{l0} \ 
{\forall}l{\in}[0,n-1].\eqno(3.13)$$
\vskip 0.2cm
(2) As ${\eta}$ increases through a certain positive value,
${\eta}_{1},$ the fixed point $x_{0}$ becomes unstable, and, by
Hopf bifurcation, gives way to a periodic orbit of the form  
$${\alpha}_{l,t}={\alpha}_{l}^{(0)}
{\exp}(-i({\nu}t+{\theta}^{(0)})){\delta}_{lk}; \
s_{l,t}=s_{l}^{(0)}{\exp}((-i{\nu}t+{\theta}^{(0)})){\delta}_{
lk};
\ p_{l,t}={\eta}_{1}{\delta}_{l0},\eqno(3.14)$$
where the selection of the $k$'th mode and the values of the
constants ${\alpha}^{(0)}, \ s^{(0)}$ and ${\nu}$ are determined
by the parameters of the model, while the phase angle
${\theta}^{(0)}$ is indeterminate. Thus, in optical terms, there
is a transition from normal to coherent radiation as ${\eta}$
passes through ${\eta}_{1}.$ This entails a breakdown of the
gauge symmetry, represented by the transformations in which the
${\sigma}_{r,-},a_{l},$ and correspondingly $s_{l},{\alpha}_{l},$
are all rephased by the same factor ${\exp}(i{\theta}).$
\vskip 0.2cm
(3) The model generically undergoes a further transition from
coherent to chaotic radiation, as ${\eta}$ passes through a
second critical value, ${\eta}_{2}(>{\eta}_{1}).$ Specifically,
the following two scenarios, both of which involve gauge symmetry
breakdown, are feasible.
\vskip 0.2cm
(a) There can be a bifurcation, at ${\eta}={\eta}_{2},$ from a
periodic orbit to a strange attractor, as specified in the
Ruelle-Takens scheme$^{(11)}.$ This scenario is actually realised
in the
single mode case$^{(4)}.$
\vskip 0.2cm
(b) For large $n,$ there could be a succession of bifurcations,
corresponding to the activation of different modes, leading to
a chaotic phase along the lines of Landau's theory of
hydrodynamical turbulence$^{(12)}$.
\vskip 0.5cm
\centerline {\bf IV. The Microscopic Dynamics in Pure Phases.} 
\vskip 0.3cm
Our objective now is to obtain the microscopic dynamics of the
model, in the limit $N{\rightarrow}{\infty}.$ We take the local
microscopic observables to be those of the matter only, since,
as noted in the Comment following Prop. 3.3, the radiation field
reduces to a classical macroscopic one in this limit, and
therefore this field carries no local microscopic observables within
the terms of this model$^{(13)}.$
\vskip 0.2cm
We formulate the $C^{\star}-$algebra of the observables of the
matter, in the limit $N{\rightarrow}{\infty},$ as the standard
quasi-local one for the infinite chain of atoms of the given
species, occupying the sites of the lattice ${\bf Z}.$ Thus, in
order to construct this algebra, we define $L$ to be the family of all
finite point subsets of ${\bf Z},$ and we assign to each
${\Lambda}{\in}L$ the $C^{\star}-$algebra
${\cal A}_{\Lambda}:={\otimes}_{r{\in}{\Lambda}}{\cal A}_{r},$
with ${\cal A}_{r}$ as defined in Section 2. For
${\Lambda}{\subset}{\Lambda}^{\prime},$ we identify elements $A$
of ${\cal A}_{\Lambda}$ with
$A{\otimes}I_{{\Lambda}^{\prime}{\backslash}{\Lambda}} \
({\in}{\cal A}_{{\Lambda}^{\prime}}),$ thereby rendering ${\cal
A}_{\Lambda}$ isotonic w.r.t. ${\Lambda}.$ We then define ${\cal
A}$ to be the completion of the local algebra ${\cal
A}_{L}:={\bigcup}_{{\Lambda}{\in}L}{\cal A}_{\Lambda}$ w.r.t.
the norm it inherits from the ${\cal A}_{\Lambda}$'s. Thus, the
algebras ${\cal A}^{(N)}$ of the material observables of
the systems ${\Sigma}^{(N)}$ are just the local subalgebras
${\cal A}_{[-N,N]}$ of ${\cal A}.$ 
\vskip 0.2cm
We denote by ${\cal S}({\cal A})$ the set of all states on ${\cal
A}$, and we represent the space translation group ${\bf Z}$
by the automorphisms $V$ of this algebra, defined by the formula 
$$V(m){\sigma}_{r,u}={\sigma}_{r+m,u}, \
{\forall}r,m{\in}{\bf Z}, \ u={\pm},z.\eqno(4.1)$$
\vskip 0.2cm
We note that it follows from our specifications of the local
structure of ${\cal A}$ that the semigroup $T_{mat}^{(N)},$
defined in Section 2, is just the restriction to ${\cal A}^{(N)}$
of the global one-parameter semigroup
$T_{mat}:={\otimes}T_{r}$ of CP transformations of
${\cal A}.$ The generator of $T_{mat}$ is therefore 
$$L_{mat}={\sum}_{r{\in}{\bf Z}}L_{r},\eqno(4.2)$$
where $L_{r}$ is given by equn. (2.6). The effects of the
matter-field coupling on the properties of ${\Sigma}^{(N)},$ in
the limit $N{\rightarrow}{\infty},$ will be treated presently.
\vskip 0.2cm
Our aim now will be to obtain the microdynamics of the matter
induced by that of the systems ${\Sigma}^{(N)}$ in the limit
$N{\rightarrow}{\infty},$ subject to the following conditions,
characteristic of pure phases, on their initial states. 
\vskip 0.2cm
(P1) The restrictions of the initial states, ${\psi}^{(N)},$ of
the systems ${\Sigma}^{(N)},$ to the algebras
${\cal A}^{(N)}{\equiv}{\cal A}_{[-N,N]}$ are the same as those
of a certain primary element, ${\omega},$ of ${\cal S}({\cal
A}),$ i.e.,
$${\omega}_{{\vert}{\cal A}^{(N)}}={\psi}^{(N)}_{{\vert}{\cal
A}^{(N)}}, \ {\forall}N{\in}{\bf N}.$$
\vskip 0.2cm
(P2) ${\omega}(s_{l}^{(N)})$ and ${\omega}(p_{l}^{(N)})$
converge to values $s_{l}$ and $p_{l},$ respectively, for each
$l{\in}[0,n-1],$ as $N{\rightarrow}{\infty}.$
\vskip 0.2cm
(P3) The mean and dispersion of ${\alpha}_{l}^{(N)},$ for the
state ${\psi}^{(N)},$ converge to values ${\alpha}_{l}$ and zero,
respectively, for each $l{\in}[0,n-1],$ as
$N{\rightarrow}{\infty}.$
\vskip 0.2cm
The following lemma serves to confirm that the assumptions (P1-
3), together with equn. (3.7), imply that the conditions of Prop.
3.3(iii) prevail.
\vskip 0.3cm
{\bf Lemma 4.1.} {\it Under the above assumptions, the
macroscopic probability measure, $m_{t},$ is simply
${\delta}_{x_{t}},$ where $x_{t}={\tau}(t)x$ and $x$ is the
initial phase point corresponding to $({\alpha},s,p)$ of
conditions (P2,3).}
\vskip 0.3cm
{\bf Proof.} By equn. (3.7) and the Schwartz inequality,
$${\vert}{\mu}_{0}^{(N)}(y)-{\exp}(iy.x){\vert}{\equiv}
{\vert}{\psi}^{(N)}\bigl({\exp}(iy.(x^{(N)}-x))-I\bigr){\vert}
{\leq}
2\bigl[{\psi}^{(N)}\bigl({\sin}^{2}({y.(x^{(N)}-x)\over
2})\bigr)\bigr]^{1/2}$$
$${\leq}\bigl[{\psi}^{(N)}\bigl(y.(x^{(N)}-x)\bigr)^{2}\bigr]^
{1/2},$$ 
and, in view of (P2,3), this tends to zero as
$N{\rightarrow}{\infty}.$ Hence, by Prop. 3.3(i),
${\mu}_{0}(y)={\exp}(iy.x),$ i.e., $m_{0}={\delta}_{x};$ and
consequently, by Prop. 3.3(iii), $m_{t}={\delta}_{x_{t}},$ as
required.
\vskip 0.3cm 
The following Proposition signifies that, in the limit
$N{\rightarrow}{\infty}$, the microscopic dynamics is precisely
that of the matter, under the action of the classical macroscopic
radiation field, ${\phi}.$ 
\vskip 0.3cm
{\bf Proposition 4.2.} {\it (i) Assuming the conditions (P1-3),
the microscopic dynamics of the model corresponds, in the limit
$N{\rightarrow}{\infty},$ to a two-parameter family,
$${\lbrace}{\cal T}(s,t{\vert}{\phi}){\vert}0{\leq}s{\leq}t;
{\cal T}(s,u{\vert}{\phi}){\cal T}_{\mu}(u,t{\vert}{\phi})=
{\cal T}(s,t{\vert}{\phi}){\rbrace},$$ 
of CP contractions of ${\cal A},$ in that 
$${\lim}_{N\to\infty}{\psi}^{(N)}\bigl(T^{(N)}(t)A\bigr)=
{\omega}\bigl({\cal T}(0,t{\vert}{\phi})A\bigr) \ {\forall}
A{\in}{\cal A}_{L},t{\in}{\bf R}_{+}.\eqno(4.3)$$
Furthermore, the generator of ${\cal T}$ is
$${\cal L}(t{\vert}{\phi}):={{\partial}\over
{\partial}t}{\cal T}(s,t{\vert}{\phi})_{{\vert}s=t}=L_{mat}+
{\sum}_{r{\in}{\bf Z}}[{\phi}_{r,t}{\sigma}_{r,+}
-h.c.,.].\eqno(4.4)$$
Thus,
$${\cal T}(s,t{\vert}{\phi})={\bf T}_{a}
{\exp}\bigl(\int_{s}^{t}du{\cal L}(u{\vert}{\phi})\bigr),
\eqno(4.5)$$
where ${\bf T}_{a}$ is the antichronological operator.
We define
$${\omega}_{t}:={\omega}{\circ}{\cal T}(0,t).\eqno(4.6)$$
\vskip 0.2cm
(ii) ${\cal T}$ factorises into a product of single site
contributions according to the formula
$${\cal T}(s,t{\vert}{\phi})={\otimes}_{r{\in}{\bf Z}}
{\cal T}_{r}(s,t{\vert}{\phi}_{r}),\eqno(4.7)$$
where
$${\cal T}_{r}(s,t{\vert}{\phi}_{r})={\bf T}_{a}
{\exp}\bigl(\int_{s}^{t}du{\cal L}_{r}(u{\vert}{\phi})\bigr),
\eqno(4.8)$$
and
$${\cal L}_{r}(t{\vert}{\phi})=L_{r}+
[{\phi}_{r,t}{\sigma}_{r,+}-h.c.,.].\eqno(4.9)$$}
\vskip 0.3cm
{\bf Comment.} This result is similar to that obtained by
Bagarello 
and Morchio$^{(9)}$ for a class of conservative systems, in that
the 
microscopic evolution is governed by a classical field, formed
by 
the time-dependent macroscopic observables.
\vskip 0.3cm
The proof of the Proposition requires the following lemmas.
\vskip 0.3cm
{\bf Lemma 4.3.} {\it Defining 
$${\tilde {\alpha}}_{l,t}^{(N)}:={\alpha}_{l}^{(N)}-
{\alpha}_{l,t},\eqno(4.10)$$
$${\lim}_{N\to\infty}{\psi}_{t}^{(N)}
({\tilde {\alpha}}_{l,t}^{(N){\star}}
{\tilde {\alpha}}_{l,t}^{(N)})=
{\lim}_{N\to\infty}{\psi}_{t}^{(N)}
({\tilde {\alpha}}_{l,t}^{(N)}
{\tilde {\alpha}}_{l,t}^{(N){\star}})=0,\eqno(4.11)$$
the convergence being uniform on the $t-$compacts.}
\vskip 0.3cm
{\bf Lemma 4.4.} {\it Defining
$${\tilde s}_{l,t}^{(N)}:=s_{l}^{(N)}-s_{l,t},\eqno(4.12)$$
$${\lim}_{N\to\infty}{\psi}_{t}^{(N)}
({\tilde s}_{l,t}^{(N)}{\tilde s}_{l,t}^{(N){\star}})=0,
\eqno(4.13)$$
the convergence being uniform on the $t$-compacts.}
\vskip 0.3cm
{\bf Proof of Prop. 4.2, assuming Lemma 4.3.} (i) By equns.
(2.5), (2.6), (2.9), (2.10), (2.12) and (2.13),
$${\psi}_{t}^{(N)}(A)-
{\psi}_{0}^{(N)}(A)=\int_{0}^{t}
ds{\psi}_{s}^{(N)}(L_{mat}A)+$$
$$\int_{0}^{t}ds{\sum}_{l=0}^{n-1}
{\psi}_{s}^{(N)}({\alpha}_{l}^{(N)}{\delta}_{l}^{+}A+
{\alpha}_{l}^{(N){\star}}{\delta}_{l}^{-}A) \
{\forall}A{\in}{\cal A}_{L}, \ t{\in}{\bf
R}_{+},\eqno(4.14)$$
where 
$${\delta}_{l}^{\pm}={\pm}{\lambda}_{l}{\sum}_{r{\in}{\bf Z}}
{\exp}({\pm}2{\pi}irl/n)[{\sigma}_{r,{\pm}},.].\eqno(4.15)$$
We note now that ${\cal A}_{L}$ is stable under
${\delta}_{l}^{\pm},$ and that, by equns. (2.14), (2.15) and
(3.3),
${\psi}_{t}^{(N)}({\alpha}_{l}^{(N){\star}}{\alpha}_{l}^{(N)})$
is bounded, uniformly w.r.t. $N$ and $t.$
It therefore follows from the Schwartz inequality that, for each
$A{\in}{\cal A}_{L},$ the norm of the r.h.s. of equn. (4.14)
is likewise uniformly bounded. Hence, by the Arzela-Ascoli
theorem, ${\psi}_{t}^{(N)}(A)$ converges to a limit, uniformly 
over the $t-$compacts, as $N{\rightarrow}{\infty}$ over some 
subsequence of ${\bf N}$. Therefore, as ${\cal A}$ is norm-separable, 
it follows from the standard diagonalisation procedure that there is 
a state ${\omega}_{t}^{\prime}$ on ${\cal A},$ such that, over some
subsequence of the integers,
$${\lim}_{N\to\infty}{\psi}_{t}^{(N)}(A)
={\omega}_{t}^{\prime}(A) \ {\forall}A{\in}{\cal A}_{L},
t{\in}{\bf R}_{+}.\eqno(4.16)$$
Further, by equns. (2.7), (4.2), (4.4), (4.8) and (4.10), we may
re-express equn. (4.14) in the form
$${\psi}_{t}^{(N)}(A)-{\psi}_{0}^{(N)}(A)=\int_{0}^{t}ds
{\psi}_{s}^{(N)}
\bigl({\cal L}(s{\vert}{\phi})A\bigr)+
\int_{0}^{t}ds{\sum}_{l=0}^{n-1}{\psi}_{s}^{(N)}
({\tilde {\alpha}}_{l,s}^{(N)}{\delta}_{l}^{+}A+
{\tilde {\alpha}}_{l,s}^{(N){\star}}
{\delta}_{l}^{-}A).\eqno(4.17)$$
By Schwartz's inequality and Lemma 4.3, the last integral in this
equation tends to zero, uniformly on the compacts, as
$N{\rightarrow}{\infty}.$ Hence, by equn. (4.16), we see that
equn. (4.17) reduces to the following form in this limit.
$${\omega}_{t}^{\prime}(A)-{\omega}_{0}^{\prime}(A)
=\int_{0}^{t}ds{\omega}_{s}^{\prime}
\bigl({\cal L}(s{\vert}{\phi})A\bigr),$$
and hence
$${d\over dt}{\omega}_{t}^{\prime}(A)
={\omega}_{t}^{\prime}\bigl({\cal L}(t{\vert}{\phi})A\bigr).
\eqno(4.18)$$
Consequently, since equn. (4.5) implies that
$${{\partial}\over {\partial}s}
{\cal T}(s,t{\vert}{\phi})=
-{\cal L}(s{\vert}{\phi})
{\cal T}(s,t{\vert}{\phi}),$$
it follows from equn. (4.18) that
$${{\partial}\over {\partial}s}
{\omega}_{s}^{\prime}({\cal T}(s,t{\vert}{\phi})A)=0 \
{\forall}0{\leq}s{\leq}t, \ A{\in}{\cal A}_{L}.$$
Hence, ${\omega}_{s}^{\prime}\bigl({\cal
T}(s,t{\vert}{\phi})A\bigr)$ is independent of $s,$ and
consequently, by assumption (P1) and equn. (4.16),
$${\omega}_{t}^{\prime}(A)=
{\omega}_{0}^{\prime}\bigl({\cal T}(0,t{\vert}{\phi})A\bigr)
{\equiv}{\omega}\bigl({\cal T}(0,t{\vert}{\phi})A\bigr),$$
i.e., by equn. (4.6),
$${\omega}_{t}^{\prime}={\omega}_{t}
={\omega}{\circ}{\cal T}(0,t{\vert}{\phi}).\eqno(4.19)$$
Equations (4.3) and (4.4) now follow, at least for convergence
over a subsequence of integers, from equns. (4.16), (4.18) and
(4.19). The removal of the restriction to a subsequence is
achieved by noting that, in view of the uniqueness of the
evolution ${\omega}{\rightarrow}{\omega}_{t},$ specified by equn.
(4.6), the compactness argument underlying the derivation of
equn. (4.19) may be re-employed to show that all subsequential limits 
of ${\lbrace}{\psi}^{(N)}(T^{(N)}(t)A){\vert}N{\in}{\bf N}{\rbrace}$
are equal to ${\omega}_{t}(A)$. 
\vskip 0.2cm
(ii) The factorisation property is an immediate consequence of
the fact that, by equns. (4.4) and (4.9),
$${\cal L}(t{\vert}{\phi})={\sum}_{r{\in}{\bf Z}}
{\cal L}_{r}(t{\vert}{\phi}_{r}).$$
\vskip 0.3cm
{\bf Proof of Lemma 4.3, assuming Lemma 4.4.} Since, by equns.
(2.7), (3.3) and (4.10),
$[{\tilde {\alpha}}_{l,t}^{(N)},
{\tilde {\alpha}}_{l,t}^{(N){\star}}]=N^{-1},$
it suffices to establish that the l.h.s. of equn. (4.11) tends
to zero, uniformly on the $t-$compacts, as
$N{\rightarrow}{\infty}.$
\vskip 0.2cm
To this end, we define 
$$J_{l,t}^{(N)}:={\psi}_{t}^{(N)}
({\tilde {\alpha}}_{l,t}^{(N){\star}}
{\tilde {\alpha}}_{l,t}^{(N)}),\eqno(4.20)$$
and then infer from this equation and equns. (2.12) and (4.10)
that
$${dJ_{l,t}^{(N)}\over dt}={\psi}_{t}^{(N)}\bigl((L^{(N)}+
{{\partial}\over
{\partial}t}){\tilde {\alpha}}_{l,t}^{(N){\star}}
{\tilde {\alpha}}_{l,t}^{(N)}\bigr).$$
It follows from this equation, together with equns. (2.13),
(3.1)-(3.3), (4.10), (4.12) and (4.20) that  
$${dJ_{l,t}^{(N)}\over dt}=-2{\kappa}_{l}J_{l,t}^{(N)}+
2{\lambda}_{l}Re\bigl({\psi}_{t}^{(N)}({\tilde
{\alpha}}_{l,t}^{(N){\star}}
{\tilde s}_{l,t}^{(N)})\bigr),\eqno(4.21)$$
and consequently, 
$$J_{l,t}^{(N)}=J_{l,0}^{(N)}{\exp}(-2{\kappa}_{l}t)+
2{\lambda}_{l}Re\int_{0}^{t}dt^{\prime}{\exp}
(-2{\kappa}_{l}(t-t^{\prime}))
{\psi}_{t^{\prime}}^{(N)}
({\tilde {\alpha}}_{l,t^{\prime}}^{(N){\star}}
{\tilde s}_{l,t^{\prime}}^{(N)}).
\eqno(4.22)$$
Therefore, defining
$$K_{l,t}^{(N)}:={\psi}_{t}^{(N)}({\tilde s}_{l,t}^{(N)}
{\tilde s}_{l,t}^{(N){\star}}),\eqno(4.23)$$
it follows from the Schwartz inequality and equn. (4.22) that
$$J_{l,t}^{(N)}{\leq}J_{l,0}^{(N)}{\exp}(-2{\kappa}_{l}t)+
\int_{0}^{t}dt^{\prime}{\vert}{\lambda}_{l}{\vert}
{\exp}(-2{\kappa}_{l}(t-t^{\prime}))[K_{l,t^{\prime}}^{(N)}
J_{l,t^{\prime}}^{(N)}]^{1/2}.\eqno(4.24)$$
For any fixed $T(>0),$ we now define
$$C_{l,T}^{(N)}:=sup_{t{\in}[0,T]}J_{l,t}^{(N)}\eqno(4.25)$$
and
$$D_{l,T}^{(N)}:=sup_{t{\in}[0,T]}K_{l,t}^{(N)},\eqno(4.26)$$
noting that these are both finite, in view of the equn. (2.15)
and the uniform boundedness of $s^{(N)}.$ Hence, by equns.
(4.24)-(4.26),
$$J_{l,t}^{(N)}{\leq}J_{l,0}^{(N)}{\exp}(-2{\kappa}_{l}t)+
2{\vert}{\lambda}_{l}{\vert}\int_{0}^{t}dt^{\prime}
{\exp}(-2{\kappa}_{l}(t-t^{\prime}))[C_{l,T}^{(N)}D_{l,T}^{(N)
}]^{1/2}
 \ {\forall} \ t{\in}[0,T].\eqno(4.27)$$
From this formula it follows easily that
$$J_{l,t}^{(N)}{\leq}J_{l,0}^{(N)}+
{{\vert}{\lambda}_{l}{\vert}\over {\kappa}_{l}}
[C_{l,T}^{(N)}D_{l,T}^{(N)}]^{1/2},$$
and hence that
$$C_{l,T}^{(N)}{\leq}J_{l,0}^{(N)}+
{{\vert}{\lambda}_{l}{\vert}\over {\kappa}_{l}}
[C_{l,T}^{(N)}D_{l,T}^{(N)}]^{1/2},\eqno(4.28)$$
which in turn implies that
$$C_{l,T}^{(N)}{\leq}{1\over 4}
\bigl[{{\vert}{\lambda}_{l}{\vert}\over
{\kappa}_{l}}(D_{l,T}^{(N)})^{1/2}+
({{\lambda}_{l}^{2}\over {\kappa}_{l}^{2}}D_{l,T}^{(N)}+
4J_{l,0}^{(N)})^{1/2})\bigr]^{2}.\eqno(4.29)$$
Further, by definition (4.20) and assumption (P3), $J_{l,0}^{(N)}$
converges to zero as $N{\rightarrow}{\infty};$ and by Lemma 4.4,
the same is true of $D_{l,T}^{(N)}.$ Hence, by equn. (4.29),
$C_{l,T}^{(N)}$ tends to zero as $N{\rightarrow}{\infty}.$ By
equns. (4.20) and (4.25), this is just the required result.
\vskip 0.3cm
{\bf Proof of Lemma 4.4.} Let $q_{l}^{(N)}$ denote the Hermitian
or anti-Hermitian part of $s_{l}^{(N)};$ and, correspondingly,
let $q_{l,t}$ denote the real or imaginary part of $s_{l,t}.$
Then, defining
$$c_{l,t}^{(N)}(v):={\psi}_{t}^{(N)}({\exp}(ivq_{l}^{(N)}))
 \ {\forall}v{\in}{\bf R},\eqno(4.30)$$   
we see from equn. (3.7) that $c_{l,t}^{(N)}$ is just the
restriction of the characteristic function ${\mu}_{t}^{(N)}$ to
one matter mode. Thus, by Prop. 3.3(iii) and Lemma 4.1,
$${\lim}_{N\to\infty}c_{l,t}^{(N)}(v)={\exp}(ivq_{t,l}), \ 
{\forall}v{\in}{\bf R},t{\in}{\bf R}_{+},\eqno(4.31)$$  
convergence being uniform on the compacts. On integrating this
equation against the Fourier transform, ${\hat f}(v),$ of any
element, $f,$ of the Schwartz space ${\cal D}({\bf R}),$ we see
that
$${\lim}_{N\to\infty}{\psi}_{t}^{(N)}(f(q_{l}^{(N)}))=f(q_{l,t})
 \ {\forall}f{\in}{\cal D}({\bf R}),t{\in}
{\bf R}_{+},\eqno(4.32)$$
convergence being uniform on the $t-$compacts for given $f.$
We shall now use this formula to prove the required result, in
the form of the equation
$${\lim}_{N\to\infty}{\psi}_{t}^{(N)}((q_{l}^{(N)})^{m})=
(q_{l,t})^{m}, \ for \ m=1,2,\eqno(4.33)$$
convergence being uniform on the compacts.
\vskip 0.2cm
To establish this formula, we first note that, by equns. (2.2)
and (3.1), together with our definition of $q_{l}^{(N)},$ the
norm of this operator is $1/2.$ Thus, if $f$ is any ${\cal
D}({\bf R})-$class function with support disjoint from
$[-1/2,1/2],$ then $f(q_{l}^{(N)})=0,$ and hence, by equn.
(4.32), $f(q_{l,t})=0.$ This implies that $q_{l,t}$ must lie in
the interval $[-1/2,1/2].$ We now re-employ equn. (4.32),
choosing $f$ to be of the form $f(q)=q^{m}h(q),$ where $m=1$ or
$2,$ and $h$ is an element of ${\cal D}({\bf R}),$ whose value
is unity over the interval $[-1/2,1/2].$ This choice yields
precisely the required result (4.33).
\vskip 0.5cm
\centerline {\bf V. Asymptotic form of the Dynamics and States
of Pure Phases}
\vskip 0.3cm
We shall now show that the microscopic dynamics of the model
simplifies greatly when transient effects, which decay in times
of the order of the lifetime of the excited states of its atoms,
are discarded.
\vskip 0.2cm
We start by noting that, by Prop. 4.2, the microscopic
dynamics of the model is completely determined by the action of
the CP contractions, ${\cal T}_{r},$ on the single site algebras
${\cal A}_{r}.$ Thus, defining  
$${\sigma}_{r,u}(t):={\cal T}_{r}(0,t){\sigma}_{r,u}, \ for \ 
u=+,-,x,y,z,\eqno(5.1)$$
we see from equations (2.1), (4.5) and (4.7)-(4.9) that 
$${d{\sigma}_{r,-}(t)\over dt}+
(i{\epsilon}+{\gamma}_{1}){\sigma}_{r,-}(t)-
i{\phi}_{r,t}{\sigma}_{r,z}(t)=0\eqno(5.2)$$
and
$${d{\sigma}_{r,z}(t)\over dt}+{\gamma}_{2}{\sigma}_{r,z}(t)
-2i({\overline {\phi}}_{r,t}{\sigma}_{r,-}(t)-h.c.)
={\gamma}_{2}{\eta}I.\eqno(5.3)$$
Thus, the spin vector
${\sigma}_{r}(t)=({\sigma}_{r,x}(t),{\sigma}_{r,y}(t),
{\sigma}_{r,z}(t))$ evolves according to a linear inhomogeneous
equation
$${d{\sigma}_{r}(t)\over dt}
=b_{r}(t){\sigma}_{r}(t)+cI,\eqno(5.4)$$
where $b_{r}(t)$ is a linear transformation of ${\bf C}^{3},$
whose explicit form is obtained from the equivalence of equn.
(5.4) to the pair of equations (5.2) and (5.3), and
$$c=(0,0,{\gamma}_{2}{\eta}).\eqno(5.5)$$
The solution of equn. (5.4) may be expressed in the form
$${\sigma}_{r}(t)=g_{r}(t,0){\sigma}_{r}(0)+
\int_{0}^{t}dt^{\prime}g_{r}(t,t^{\prime})cI,\eqno(5.6)$$
where $g_{r}$ is the Green function for $b_{r},$ determined by
the formula
$${{\partial}\over {\partial}t}g_{r}(t,t^{\prime})
=b_{r}(t)g_{r}(t,t^{\prime}) \ 
{\forall}t{\geq}t^{\prime}{\geq}0; \
g_{r}(t,t)=I_{3},\eqno(5.7)$$
and $I_{3}$ is the identity operator in ${\bf C}^{3}$.
\vskip 0.3cm
{\bf Lemma 5.1.} {\it For fixed, $t^{\prime}, \ g(t,t^{\prime})$
decays to zero, with increasing $t,$ at least as fast as
${\exp}(-{\gamma}t),$ where
${\gamma}=min({\gamma}_{1},{\gamma}_{2}).$}
\vskip 0.3cm
{\bf Proof.} Let
${\xi}_{r}$ be an arbitrary element of ${\bf R}^{3},$ and, for 
fixed $t^{\prime}({\in}{\bf R}_{+})$ and any $t>t^{\prime},$ let
${\xi}_{r}(t)=({\xi}_{r,x}(t),{\xi}_{r,y}(t),{\xi}_{r,z}(t))
:=g_{r}(t,t^{\prime}){\xi}_{r}.$ Then, by equn. (5.7), 
${\xi}_{r}(t)$ satisfies the homogeneous linear equation obtained
by replacing ${\sigma}_{r}$ by ${\xi}_{r}$ in equn. (5.4) and 
discarding the term $cI$ on the r.h.s. Hence, defining
${\xi}_{r,{\pm}}=1/2({\xi}_{r,x}{\pm}i{\xi}_{r,y}),$ it follows
from the equivalence between equn. (5.4) and the pair of
equations (5.2) and (5.3) that the equation of motion for
${\xi}_{r}(t)$ is also given by replacing ${\sigma}$ by ${\xi}$
in the latter equations and discarding the r.h.s. of (5.3). Thus,
$${d\over dt}{\xi}_{r,-}(t)+
({\gamma}_{1}-i{\epsilon}){\xi}_{r,-}(t)-
i{\phi}_{r,t}{\xi}_{r,z}(t)=0$$
and
$${d\over dt}{\xi}_{r,z}(t)+{\gamma}_{2}{\xi}_{r,z}(t)
-2i({\overline {\phi}}_{r,t}{\xi}_{r,-}(t)-c.c.)=0.$$
Hence,
$${d\over dt}{\vert}{\xi}_{r}(t){\vert}^{2}{\equiv}
{d\over dt}(4{\vert}{\xi}_{r,-}(t){\vert}^{2}+
{\xi}_{r,z}(t)^{2})=
-8{\gamma}_{1}{\vert}{\xi}_{r,-}(t){\vert}^{2}-
2{\gamma}_{2}{\xi}_{r,z}(t)^{2}=$$
$$-2[{\gamma}_{1}{\xi}_{r,x}(t)^{2}+
{\gamma}_{1}{\xi}_{r,y}(t)^{2}
+{\gamma}_{2}{\xi}_{r,z}(t)^{2}]{\leq}
-2min({\gamma}_{1},{\gamma}_{2}){\vert}{\xi}_{r}(t){\vert}^{2}.$$
Consequently, defining
${\gamma}:=min({\gamma}_{1},{\gamma}_{2}),$ 
$${d\over dt}({\vert}{\xi}_{r}(t){\vert}^{2}{\exp}(2{\gamma}t))
{\leq}0,$$
which implies that ${\vert}{\xi}_{r}(t){\vert},$ and hence
$g_{r}(t,t^{\prime}),$ decay to zero at least as fast
as ${\exp}(-{\gamma}t),$ with increasing t. 
\vskip 0.3cm
{\bf Comment.} This lemma signifies that the first term on the
r.h.s. of equn. (5.6) is merely a transient, which decays in a
time, ${\gamma}^{-1},$ of the order of the lifetime of an atomic 
excited state. Consequently, by (5.5), the asymptotic form for
${\sigma}_{r}(t),$ obtained by discarding the transients, is
$${\sigma}_{r}(t)={\theta}_{r}(t)I,\eqno(5.8)$$
where
$${\theta}_{r}(t)=\int_{0}^{t}dt^{\prime}g_{r}(t,t^{\prime})
(0,0,{\gamma}_{2}{\eta}).\eqno(5.9)$$
Hence, since the state ${\omega}_{t}$ is given by its action
on the finite monomials in the components of the
${\sigma}_{r}$'s, the following Proposition is a simple
consequence of these last two equations, together with Prop. 4.1.
\vskip 0.3cm
{\bf Proposition 5.2.} {\it The asymptotic form of the time-
dependent state ${\omega}_{t}$ carries no
correlations between the observables at the different sites and
is therefore of the form 
$${\omega}_{t}={\otimes}_{r{\in}{\bf Z}}{\omega}_{r,t},
\eqno(5.10)$$
where the atomic state ${\omega}_{r,t}$ is given by}
$${\omega}_{r,t}({\sigma}_{r})={\theta}_{r}(t).\eqno(5.11)$$
\vskip 0.3cm
{\bf Comment.} It follows immediately from this Proposition that
the asymptotic state of the model is completely determined by the
form of ${\theta}_{r}.$ We shall denote the components of
${\theta}_{r}$ analogously with those of ${\sigma}_{r}:$ thus
${\theta}_{r}=({\theta}_{r,x},{\theta}_{r,y},{\theta}_{r,z})$ and
${\theta}_{r,{\pm}}={1\over
2}({\theta}_{r,x}{\pm}{\theta}_{r,y}).$ The following Proposition
provides an explicit formula for ${\theta}_{r}$ in terms of the
asymptotic forms of the macroscopic variables $s$ and $p.$ 
\vskip 0.3cm
{\bf Proposition 5.3.} {\it The function ${\theta}_{r}$ is given
by the equations
$${\theta}_{r,-}(t)={\sum}_{l=0}^{n-1}s_{l,t}{\exp}(2{\pi}irl/n)
\eqno(5.12)$$
and}
$${\theta}_{r,z}(t)={\sum}_{l=0}^{n-1}p_{l,t}{\exp}(2{\pi}irl/n).
\eqno(5.13)$$
\vskip 0.3cm
{\bf Proof.} By equns. (5.2), (5.3), (5.7) and (5.9), the
equations of motion for ${\theta}_{r}$ are formally the same as
those for ${\sigma}_{r},$ i.e., 
$${d{\theta}_{r,-}(t)\over dt}+
(i{\epsilon}+{\gamma}_{1}){\theta}_{r,-}(t)-
i{\phi}_{r,t}{\theta}_{r,z}(t)=0\eqno(5.14)$$
and
$${d{\theta}_{r,z}(t)\over dt}+{\gamma}_{2}{\theta}_{r,z}(t)
-2i({\overline {\phi}}_{r,t}{\theta}_{r,-}(t)-c.c.)
={\gamma}_{2}{\eta}.\eqno(5.15)$$
Further, defining
$${\hat s}_{r}(t)={\sum}_{l=0}^{n-1}s_{l,t}{\exp}(2{\pi}irl/n)
\eqno(5.16)$$
and
$${\hat p}_{r}(t)={\sum}_{l=0}^{n-1}p_{l,t}{\exp}(2{\pi}irl/n),
\eqno(5.17)$$
we see, by elementary Fourier analysis, that equns. (3.9b) and (3.9c) 
are equivalent to
$${d{\hat s}_{r}(t)\over dt}+
(i{\epsilon}+{\gamma}_{1}){\hat s}_{r}(t)-
i{\phi}_{r,t}{\hat p}_{r}(t)=0\eqno(5.18)$$
and
$${d{\hat p}_{r}(t)\over dt}+{\gamma}_{2}{\hat p}_{r}(t)
-2i({\overline {\phi}}_{r,t}{\hat s}_{r}(t)-c.c.)
={\gamma}_{2}{\eta},\eqno(5.19)$$
respectively. Hence, the equations of motion (5.14) and (5.15)
for ${\theta}_{r,-}$ and ${\theta}_{r,z}$ are equivalent to those
given by (5.18) and (5.19) for ${\hat s}_{r}$ and ${\hat p}_{r}.$
Consequently, in the asymptotic regime, where the initial
conditions are irrelevant, ${\theta}_{r,-}={\hat s}_{r}$ and
${\theta}_{r,z}={\hat p}_{r}.$ By equations (5.16) and
(5.17), this is the required result.
\vskip 0.3cm
The following Corollary is an immediate consequence of Props. 5.2
and 5.3.
\vskip 0.3cm
{\bf Corollary 5.4.} {\it The asymptotic state ${\omega}_{t}$
is spatially periodic, with periodicity $n,$ i.e.,
$${\omega}_{t}={\omega}_{t}{\circ}V(n),\eqno(5.20)$$
where $V$ is the group of spatial translational automorphisms of
${\cal A}$ defined by equn. (4.1).}
\vskip 0.3cm
{\bf Explicit Asymptotic Form of ${\omega}_{t}.$} Equations
(5.10)-(5.13) serve to express the time-dependent microstate
${\omega}_{t}$ in terms of the macroscopic variables. We shall
now present its explicit form for both the normal and coherent
phases. We have no means of doing the same thing for the chaotic
phase, since we have no explicit solution for the macroscopic
dynamics there.
\vskip 0.2cm
{\bf 1. The Normal Phase.} Here, by equns. (3.13) and (5.10)-
(5.13), ${\omega}_{t}$ takes the stationary asymptotic value
${\omega}_{\infty},$ defined by the equations
$${\omega}_{\infty}={\otimes}_{r{\in}{\bf Z}}
{\omega}_{{\infty},r}\eqno(5.21a)$$
and
$${\omega}_{{\infty},r}({\sigma}_{r,{\pm}})=0; \
{\omega}_{{\infty},r}({\sigma}_{r,z})={\eta}.
\eqno(5.21b)$$
\vskip 0.2cm
{\bf 2. The Coherent Phase.} In this phase, it follows from
equns. (3.14) and (5.10)-(5.13) that the asymptotic form,
${\omega}_{t}^{coh},$ of ${\omega}_{t}$ is given by the formula
$${\omega}_{t}^{coh}=
{\otimes}_{r{\in}{\bf Z}}{\omega}_{r,t}^{coh},\eqno(5.22a)$$
where
$${\omega}_{r,t}^{coh}({\sigma}_{r,-})=
s_{l}^{(0)}{\exp}
\bigl(i({2{\pi}ikr\over n}-{\nu}t-{\theta}_{0})\bigr); \
{\omega}_{r,t}^{coh}({\sigma}_{r,z})={\eta}.\eqno(5.22b)$$
\vskip 0.2cm
{\bf Comment.} By contrast with the thermal equilibrium states
of pure phases, the stable states ${\omega}_{t}^{coh}$ are
time-dependent. The same is true for the pure chaotic phases, since 
the macrosopic variables $s_{t}$ and $p_{t}$ are time-dependent there 
too$^{(14)}$.
\vskip 0.5cm
\centerline {\bf VI. A Non-Equilibrium Maximum Entropy Principle}
\vskip 0.3cm
We shall now establish that, as a further consequence of the
results of Section 5, ${\omega}_{t}$ maximises the global
entropy density of the matter, as defined on the states
with spatial periodicity $n,$ subject to the condition that the
limiting value, as $N{\rightarrow}{\infty},$ of the expectation
values of the macroscopic variables $s^{({N})},p^{(N)}$ are
$s_{t},p_{t},$ respectively.
\vskip 0.2cm
For this purpose, we start by defining ${\cal S}_{n}$ to be the
set of spatially periodic states, ${\psi},$ on ${\cal
A},$ with period $n,$ and ${\cal S}_{n,x_{t}}$ to be the
subset of these states for which
$${\lim}_{N\to\infty}{\psi}(s_{l}^{(N)})=s_{l,t} \  
and \ {\lim}_{N\to\infty}{\psi}(p_{l}^{(N)})=p_{l,t}, \ {\forall}
l{\in}[0,n],t{\in}{\bf R}_{+}.$$
Equivalently, by equns. (3.1) and (3.2), ${\cal S}_{n,x_{t}}$
consists of the states, ${\psi},$ of spatial periodicity $n,$
which satisfy the conditions
$$n^{-1}{\sum}_{r=0}^{n-1}{\psi}({\sigma}_{r,-})
{\exp}(-2{\pi}ilr/n)=s_{l,t}\eqno(6.1)$$
and
$$n^{-1}{\sum}_{r=0}^{n-1}{\psi}({\sigma}_{r,z})
{\exp}(-2{\pi}ilr/n)=p_{l,t},\eqno(6.2)$$
for $l=0,. \ .,n-1$ and $t{\in}{\bf R}_{+}.$
\vskip 0.2cm
We formulate the global entropy density functional, $s$ on
${\cal S}_{n}$ in the following standard way$^{(15)}$. For
${\psi}{\in}{\cal S}_{n}$ and ${\Lambda}{\in}L,$ we define
${\rho}_{\Lambda}^{\psi}$ to be the density matrix in ${\cal
H}({\Lambda})$ representing the restriction of ${\psi}$ to
${\cal A}({\Lambda}),$ and $S_{\Lambda}({\psi})$ to be the
corresponding local entropy, i.e.,
$$S_{\Lambda}({\psi})=
-kTr({\rho}_{\Lambda}^{\psi}ln{\rho}_{\Lambda}^{\psi}),
\eqno(6.3)$$
$k$ being Boltzmann's constant. We define the global entropy
density functional, $s,$ on ${\cal S}_{n}$ by the formula
$$s({\psi})={\lim}_{{\Lambda}{\uparrow}{\bf Z}}
{S_{\Lambda}({\psi})\over {\vert}{\Lambda}{\vert}},\eqno(6.4)$$
the convergence being guaranteed by the subadditivity of
entropy$^{(16)}$ and the periodicity of ${\psi}.$
\vskip 0.3cm
{\bf Proposition 6.1.} {\it The asymptotic time-dependent state
${\omega}_{t}$ maximises the functional $s,$ as restricted to
${\cal S}_{n,x_{t}}.$}
\vskip 0.3cm
{\bf Proof.} We first note that, by Cor. 5.4,
${\omega}_{t}{\in}{\cal S}_{n},$ and that, by equations
(5.10)-(5.13), it satisfies the conditions (6.1) and (6.2).
Hence, ${\omega}_{t}{\in}{\cal S}_{n,x_{t}}.$
\vskip 0.2cm
Next, we observe that, if ${\psi}$ is an arbitrary element of
$S_{n,x_{t}},$ then it follows by elementary Fourier analysis
from equns. (5.12), (5.13), (6.1) and (6.2) that
${\psi}({\sigma}_{r})={\theta}_{r}(t),$ for
$r{\in}[0,n-1],$ and consequently, by the periodicity of
${\psi},$ for all $r{\in}{\bf Z}.$ Therefore, by equn. (5.11), 
the restrictions of ${\psi}$ and ${\omega}_{t}$ to each
single site algebra ${\cal A}_{r}$ are identical. In other words,
{\it all the elements of ${\cal S}_{n,x_{t}}$ coincide on each
of the single site algebras.}
\vskip 0.2cm
Further, by the subadditivity of entropy,
$$S_{\Lambda}({\psi}){\leq}
{\sum}_{r{\in}{\Lambda}}S_{{\lbrace}r{\rbrace}}({\psi}), \
{\forall}{\psi}{\in}{\cal S}_{n,x_{t}},$$
while, by equns. (5.10) and (6.3),
$$S_{\Lambda}({\omega}_{t})=
{\sum}_{r{\in}{\Lambda}}
S_{{\lbrace}r{\rbrace}}({\omega}_{t}).$$
Consequently, as ${\psi}$ and ${\omega}_{t}$ coincide on the
single site algebras, 
$$S_{\Lambda}({\psi}){\leq}S_{\Lambda}({\omega}_{t}),$$
and therefore, by equn. (6.4),
$$s({\psi}){\leq}s({\omega}_{t}) \ {\forall}{\psi}{\in}
{\cal S}_{n,x_{t}},$$
which proves the Proposition. 
\vskip 0.5cm
\centerline {\bf VII. Concluding Remarks.}
\vskip 0.3cm
We have provided a microscopic picture of the multi-mode laser
model, that is complementary to the macroscopic one of (I). Our
principal new results are the following.
\vskip 0.2cm
(1) The microscopic dynamics is governed by the classical,
macroscopic field, ${\phi},$ which in turn is generated by the
collective action of the atoms on the radiation (Prop. 4.1).
\vskip 0.2cm
(2) This microdynamics serves to drive the system into states in
which the observables of different atoms are uncorrelated (Prop.
5.2). These states have particularly simple forms in both the
normal and coherent phases, being stationary and
space-translationally invariant in the former and periodic w.r.t.
both space and time, in the latter.
\vskip 0.2cm
(3) In each phase, the eventual time-dependent
microstate of the system maximises the specific entropy of the
model, subject to the constraints imposed by the instantaneous
values of the macroscopic variables. To the best of our
knowledge, this is the first example where such a variational
principle has been established in a regime far from thermal
equilibrium.
\vskip 0.2cm
These results concern (1) an interplay between microphysics and
macrophysics, (2) the decorrelation of certain micro-observables
in the course of time, and (3) a maximum entropy principle,
corresponding to a conditional thermodynamic stability far from
thermal equilibrium. Although the model treated here is but a
caricature of a laser, one might envisage that counterparts of
these results might have at least some measure of validity
in the physics of real dissipative systems. For example, one
might expect a local version of the maximum entropy principle (3)
to prevail in hydrodynamics, in such a way as to impose local
equilibrium conditions on the states of fluids. 
\vskip 0.5cm
\centerline {\bf Acknowledgments.}
\vskip 0.3cm
This work was carried out during a visit of F. B. to Queen Mary
and Westfield College (QMW). He would like to thank the CNR
for the financial support that  enabled him to make this
visit, and to express his gratitude to the staff of QMW for their
hospitality.   
\vskip 0.5cm
\centerline {\bf References.}
\vskip 0.2cm\noindent
$^{a)}$ Permanent address: Dipartimento di Matematica ed
Applicazioni, 
Fac. Ingegneria, Universita' di Palermo, I-90128 Palermo, Italy
\vskip 0.1cm\noindent
E-mail: Bagarello@ipamat.math.unipa.it
\vskip 0.2cm\noindent
$^{b)}$ E-mail: G.L.Sewell@qmw.ac.uk
\vskip 0.2cm\noindent
$^{(1)}$ K. Hepp and E. H. Lieb: Helv. Phys. Acta {\bf 46}, 573,
1973; 
and Pp. 178-208 of "Dynamical Systems, Theory and Applications",
Springer 
Lecture Notes in Physics {\bf 38}, Ed. J. Moser, Springer,
Heidelberg, 
New York, 1975 
\vskip 0.2cm\noindent
$^{(2)}$ R. H. Dicke: Phys. Rev. {\bf 93}, 99, 1954
\vskip 0.2cm\noindent
$^{(3)}$ R. Graham and H. Haken: Z. Phys. {\bf 237}, 31, 1970; 
and H. Haken: Handbuch der Physik, Bd. XXV/2C, Springer,
Heidelberg, Berlin, New York, 1970
\vskip 0.2cm\noindent
$^{(4)}$ G. Alli and G. L. Sewell: J. Math. Phys. {\bf 36}, 5598,
1995
\vskip 0.2cm\noindent
$^{(5)}$ Apart from the difference in the their mathematical
settings, this model differs from that of HL in the following
physical respect. The atoms of HL are pairs of Fermi oscillators,
and can thus accommodate $0,1$ or $2$ electrons each; whereas
those of (I) correspond to two-level atoms, each with precisely
one electron.
\vskip 0.2cm\noindent
$^{(6)}$ G. Lindblad: Commun. Math. Phys. {\bf 48}, 119, 1976
\vskip 0.2cm\noindent
$^{(7)}$ G. Morchio and F. Strocchi: J. Math. Phys. {\bf 28},
1912, 1987
\vskip 0.2cm\noindent
$^{(8)}$ R. Haag and D. Kastler: J. Math. Phys. {\bf 5}, 848,
1964
\vskip 0.2cm\noindent
$^{(9)}$ F. Bagarello and G. Morchio: J. Stat. Phys. {\bf 66},
859, 1992
\vskip 0.2cm\noindent 
$^{(10)}$ P. Vanheuverzwijn: Ann. Inst. H. Poincare A {\bf 29},
123,
1978; Erratum {\it ibid} {\bf 30}, 83, 1979; and B. de Moen, 
P. Vanheuverzwijn and A. Verbeure: Rep. Math. Phys. {\bf 15}, 27,
1979
\vskip 0.2cm\noindent
$^{(11)}$ D. Ruelle and F. Takens: Commun. Math. Phys. {\bf 20},
167,
1971
\vskip 0.2cm\noindent
$^{(12)}$ L. D. Landau and E. M. Lifshitz: "Fluid Mechanics",
Pergamon, Oxford, New York, Toronto, Sydney, Paris, 1984 
\vskip 0.2cm\noindent
$^{(13)}$ In the present model, this is due to the fact that the
radiation field is truncated to $n$ modes. Note that here, as in
the non-truncated model of the radiation, the Fourier
coefficients of ${\phi}^{(N)}$ are $N-$independent multiples of
the ${\alpha}^{(N)}$'s, rather than the $a$'s; and consequently
the latter do not appear in the local field algebra when one
passes to the limit $N{\rightarrow}{\infty}.$
\vskip 0.2cm\noindent
$^{(14)}$ If $s_{t}$ and $p_{t}$ were constant, then it 
would follow from equn. (3.9a) that ${\alpha}_{t}$, and hence $x_{t}$, 
would tend to a constant as $t$ tended to infinity, and consequently 
that the macroscopic dynamics would not be chaotic.
\vskip 0.2cm\noindent
$^{(15)}$ D. Ruelle: "Statistical Mechanics", W. A. Benjamin,
Inc., New York, 1969
\vskip 0.2cm\noindent
$^{(16)}$ E. H. Lieb and M. B. Ruskai: J. Math. Phys. {\bf 14},
1938, 1973
\end